\documentstyle[aps,manuscript]{revtex}  
\makeatletter
\newcounter{subeqncnt}
\def\thesubeqncnt{\alph{subeqncnt}}
\def\subequations{\begingroup%
   \stepcounter{equation}\edef\@tempa{\theequation}%
   \let\c@equation\c@subeqncnt\c@subeqncnt\z@
   \edef\theequation{\@tempa\noexpand\thesubeqncnt}}

\makeatother

\begin{document}                
\title{Efficient Recursion Method for Inverting Overlap Matrix}
\author{T. Ozaki}
\address{
     RICS,
     National Institute of Advanced Industrial Science and Technology (AIST),
     central 2, 1-1-1 Umezono, Tsukuba, Ibaraki 305-8568, Japan
     and
     JRCAT-ATP,
     central 4, 1-1-1 Higashi, Tsukuba, 
     Ibaraki 305-0046, Japan
}
\maketitle
\begin{abstract}                
     A new O($N$) algorithm based on a recursion method, in which the
     computational effort is proportional to the number of atoms $N$,
     is presented for calculating the inverse of an overlap matrix which
     is needed in electronic structure calculations with the the
     non-orthogonal localized basis set. This efficient inverting
     method can be incorporated in several O($N$) methods
     for diagonalization of a generalized secular equation.
     By studying convergence properties of the 1-norm of an error matrix
     for diamond and fcc Al, this method is compared to three other O($N$)
     methods (the divide method, Taylor expansion method, and Hotelling's
     method) with regard to computational accuracy and efficiency within
     the density functional theory.
     The test calculations show that the new method
     is about one-hundred times faster than the divide method in
     computational time to achieve the same convergence for both diamond
     and fcc Al, while the Taylor expansion method and Hotelling's method
     suffer from numerical instabilities in most cases.
\end{abstract}
\vspace{2cm}

 The development of O($N$) methods
 \cite{Pettifor,Ozaki,Goedecker,Stephan,Yang,Galli,Mauri,Daw,Li,Palser}
 and the revival of localized
 orbitals as a basis set
 \cite{Sankey,Kobayashi,Kurita,Kobayashi2,Hierse,Hernandez,Ordejon,Sanchez,Horsfield}
 have been made during the last decade
 in order to extend the applicability of the first-principles molecular
 dynamics (FPMD) simulations using the plane wave expansion and the
 Car-Parrinello method within density functional
 theories (DFT) \cite{Payne}.
 However, only few applications of these ${\rm O}(N)$ methods to large
 systems have been reported within the DFT calculations
 \cite{Sanchez,Bowler,Applications}.
 Although there are a lot of limitations of the method based on
 the localized description \cite{Applications}, one of the limitations
 is that several O($N$) methods require evaluating the inverse
 of the overlap matrix $S$ which comes from non-orthogonality among
 the localized orbitals.

 In the generalized Fermi operator expansion (FOE) method \cite{Stephan}
 to the non-orthogonal basis we need to calculate the inverse of
 overlap matrix to construct the modified Hamiltonian $H'\equiv S^{-1}H$,
 while Stephan et al. have proposed solving a linear equation $SH'=H$
 with the cutoff radii of $H$ instead of calculating the inverse of
 overlap matrix.
 In the density matrix (DM) method \cite{Daw,Li,Palser} which is
 a promising approach for materials with a wide gap, fortunately,
 the evaluation of the inverse is not required during the optimization
 of grand potentials, although we have to evaluate the inverse of the
 overlap matrix for a good initial guess of the density matrix \cite{Palser}.
 The block bond-order potential (BOP) method \cite{Ozaki}, which has good
 convergence properties for both insulators and metals, also
 requires the evaluation of the modified Hamiltonian $H'$ as in method
 the FOE method. If the overlap matrix is sparse, the computational
 cost scales as the second power of the number of atoms $N$ in the
 inverse calculation. Therefore, an efficient O($N$) method
 for inverting the overlap matrix should be developed.

 So far, several O($N$) inverting methods have been proposed.
 Gibson et al. used a simple method in which a linear equation
 $SH'=H$ constructed for a finite cluster is solved without
 explicit calculation of $S^{-1}$ \cite{Gibson}.
 Mauri et al. considered approximating the inverse of
 overlap matrix by the Taylor expansion \cite{Mauri}. The approach could be 
 an O($N$) inverting method when the matrix elements in the $p$th
 moment $O^p$ of the overlap matrix $O$ are cut at a finite distance.
 Palser and Manolopoulos proposed to evaluate the inverse 
 by Hotelling's method which is similar to the iterative
 purification algorithm of the DM method \cite{Palser}.
 The iterative calculation can be performed in O($N$) operations,
 provided that the cutoff of matrix elements at a finite distance is
 introduced in the product of two matrices.
 It is worth pointing out that the ideas of these
 O($N$) inverting methods are analogous to those of the
 O($N$) methods for the diagonalization. 
 The divide method by Gibson et al. \cite{Gibson}, the Taylor expansion
 method \cite{Mauri}, and Hotelling's method \cite{Palser} strategically
 and mathematically correspond to the divide and conquer method \cite{Yang},
 the FOE method \cite{Goedecker,Stephan}, and the DM method
 \cite{Daw,Li,Palser}, respectively.
 Therefore, one may expect that these O($N$) inverting methods
 may have the convergence properties for realistic materials
 similar to the O($N$) methods for the diagonalization \cite{Comparison}.
 However, it remains to be seen whether the expectation is meaningful
 or not.

 In this paper we propose a new O($N$) method for calculating 
 the inverse of the overlap matrix which is based on a resolvent and
 the block Lanczos algorithm. The new method is compared 
 with the other three methods in terms of the computational accuracy
 and efficiency. Thus, our aim of this paper is to clarify the
 applicability of these four O($N$) inverting methods for
 realistic materials.
 The paper is organized as follows. In Sec. II we present the theory
 of a new O($N$) inverting method based on a recursion method,
 and also summarize the three other O($N$) inverting methods.
 In Sec. III we discuss the convergence properties of these four
 O($N$) inverting methods for the diamond and fcc Al within
 the DFT calculations using the 1-norm of an error matrix
 which will be related to the error in the eigenvalues in this section.
 In Sec. IV we conclude with clear characterization of the
 four O($N$) inverse methods.

 \begin{center}
   {\bf II.~THEORY}
 \end{center}

 \begin{center}
   {\bf A. Recursion method}
 \end{center}

 It is assumed that one-particle wave functions are expanded 
 by a localized orbital basis set $(\vert i\alpha\rangle)$, where 
 $i$ is a site index and $\alpha$ is an orbital index.
 The localized orbitals could be Slater-type
 \cite{Kobayashi,Kurita,Kobayashi2}, Gaussian-type \cite{Hierse},
 and numerical orbitals \cite{Sankey,Hernandez} obtained by
 DFT calculations for atoms.
 In most cases, the orbitals are non-orthogonal between them,
 leading to an overlap matrix $S$ defined by 
 \begin{eqnarray}
   S_{i\alpha,j\beta} = \langle i\alpha \vert \hat{S}\vert j\beta \rangle,
 \end{eqnarray}
 where $\hat{S}$ is the overlap operator which is introduced as a matter
 of form in order to emphasize the similarity
 between the new inverting method and the block BOP method \cite{Ozaki},
 although the overlap operator generally should be the identity operator I.
 The overlap integral exponentially decays in real space 
 because of the localized nature of the orbitals, so that 
 the overlap matrix $S$ is sparse. Here we introduce a resolvent
 $R(Z)$ for the matrix $S$ as follows:
 \begin{eqnarray}
   R(Z) = (S-Z{\rm I})^{-1}.
 \end{eqnarray}
 It is then easy to verify that  
 \begin{eqnarray}
   S^{-1} = {\rm Re}R(0).
 \end{eqnarray}
 Thus, we see that the real part of the resolvent for $Z=0$
 gives the inverse $S^{-1}$ of the overlap matrix. 
 If the resolvent for $Z=0$ has a finite value for the imaginary part,
 the basis set is not linearly independent.
 The resolvent can be evaluated by adopting the algorithm of the
 block BOP method \cite{Ozaki} which is recently developed to simulate
 orthogonal tight-binding (TB) models in O($N$) operations.
 It is noted that the new inverting method is derived just by replacing
 the Hamiltonian $\hat{H}$ in the block BOP method within the orthogonal
 TB models with the overlap operator $\hat{S}$.
 The first step in this algorithm is to block-tridiagonalize 
 the overlap matrix $S$ using the block Lanczos algorithm
 \cite{Lanczos,Jones,Inoue,Haydock}.
 The central equations is 
 \begin{eqnarray}
   \hat{S}\vert U_{n}) & = & \vert U_{n})\underline{A}_{n}
                       +
            \vert U_{n-1})\underline{B}_{n}
                       +
            \vert U_{n+1})\underline{B}_{n+1}
 \end{eqnarray}
 with 
 \begin{eqnarray}
    \vert U_0) =
            (\vert i1\rangle,\vert i2\rangle,\dots,\vert iM_i\rangle )
 \end{eqnarray}
 as the starting state. $\underline{A}_n$ and $\underline{B}_n$ are 
 recursion block coefficients with $M_{i}\times M_{i}$ in size, 
 where $M_{i}$ is the number of localized orbitals on the starting 
 atom $i$, and the underline indicates that the element is a block.
 In the block Lanczos algorithm, we need to start the recursion with
 Eq.~(5) to make the recursion method accurate and efficient \cite{Ozaki}.
 The Lanczos algorithm with a finite recursion transforms the overlap
 matrix $S$ into the block-tridiagonalized matrix $S^L$ which has
 the diagonal $A_{n}$ and the sub-diagonal block elements $B_{n}$,
 where the index $L$ indicates the representation based on the Lanczos
 basis. Considering the resolvent $R^{L}(Z)\equiv (S^{L}-Z{\rm I})^{-1}$
 for the block-tridiagonalized overlap matrix,
 the diagonal $\underline{R}^L_{00}(Z)$ and off-diagonal block elements 
 $\underline{R}^L_{0n}(Z)$ can be easily derived along the same line
 as that described in the block BOP method \cite{Ozaki}.
 For $Z=0$, the elements are given by 
 \begin{eqnarray}
   \underline{R}^L_{00}(0)
        =[\underline{A}_0-\hspace{0.4mm}^t\hspace{-0.4mm}\underline{B}_1[
          \underline{A}_1-\hspace{0.4mm}^t\hspace{-0.4mm}\underline{B}_2[
                       \cdots
          ]^{-1}\underline{B}_2
          ]^{-1}\underline{B}_1
          ]^{-1},
 \end{eqnarray}
 \begin{eqnarray}
    \nonumber
    \lefteqn{ 
      \underline{R}^{L}_{0n}(0)
      =
    \biggl(
      \delta_{1n}\underline{\rm I}
      -\underline{R}^{L}_{0n-1}(0)\underline{A}_{n-1}
    }\\
    &&
    \quad\quad\quad\quad
       -\underline{R}^{L}_{0n-2}(0)
         \hspace{0.4mm}^t\hspace{-0.4mm}\underline{B}_{n-1}
    \biggr)
         (\underline{B}_{n})^{-1},
 \end{eqnarray}
 where $\delta$ is Kronecker's delta, and 
 $R_{0-1}(0)=\hspace{0.4mm}^t\hspace{-0.4mm}B_{0}=0$.
 Once the block diagonal element is calculated as the multiple
 inverse Eq.~(6), the off-diagonal elements are evaluated
 from the recurrence relation Eq.~(7) with $\underline{R}^L_{00}(0)$
 as the starting element. In order to truncate the multiple inverse
 in Eq.~(6) without reducing the accuracy significantly, a square root
 terminator could be used, while there could
 be an infinite number of levels in the multiple inverse of diagonal
 Green's function for an infinite system.
 In the test calculations of Sec.~III we used the square root
 temninator for the truncation at a finite number of levels.
 The two Eqs.~(6) and (7) provide the resolvent based on the Lanczos
 basis representation, so that we can obtain the original resolvent
 through the following inverse transformation:
 \begin{eqnarray}
    \underline{R}_{ij}(0) = \sum_{n}
                 \underline{R}^L_{0n}(0) 
                 \hspace{0.4mm}^t\hspace{-0.4mm}\underline{U}_{nj},
 \end{eqnarray}
 where $\hspace{0.4mm}^t\hspace{-0.4mm}\underline{U}_{nj}$ is defined by
 $\hspace{0.4mm}^t\hspace{-0.4mm}\underline{U}_{nj} = (U_{n}\vert
        (\vert j1\rangle,\vert j2\rangle,\dots,\vert jM_j\rangle ).$
 The inverse transformation Eq.~(8) is significantly simplified 
 because of the orthogonality in the Lanczos bases. Therefore, we only
 have to evaluate the 0th block line of the resolvent in the Lanczos
 basis representation.
 The resolvent exactly satisfies a sum rule $\sum_{ij}
  {\rm tr\left\{\underline{S}_{ij}\underline{R}_{ji}(0)\right\}}
  = N_{B}$ which is derived from Eq.~(2), where $N_{B}$ is the 
 number of bases, and is constructed by up to (q+1)th moments
 $S^{q+1}$ \cite{Ozaki}, where $q$ is a final level for the recursion.
 Equation (8) gives a good approximation for the inverse of
 overlap matrix as the number of recursion levels increases.
 However, the approximated inverse is not strictly
 a symmetric matrix at a finite recursion.
 If the approximated inverse is symmetric, eigenvalues of
 a generalized secular equation with the overlap matrix
 are real numbers. Therefore, we evaluate the inverse of
 overlap matrix by symmetrizing the resolvent in terms of
 simple arithmetic average:
 \begin{eqnarray}
    \underline{S}^{-1}_{ij} = 
    \frac{{\rm Re}\underline{R}_{ij}(0)
        + {\rm Re}\hspace{0.4mm}^t\hspace{-0.4mm}\underline{R}_{ji}(0)}
         {2}.
 \end{eqnarray}
 The symmetrization preserves the above sum rule.
 The all elements of the inverse are evaluated by applying the
 series of the algorithm repeatedly to each atom.
 The cluster over which the hops are made in the Lanczos algorithm is
 determined by the logical truncation method \cite{Ozaki}.
 Thus, the computational cost of the recursion method is strictly
 proportional to the number of atoms $N$.

 \begin{center}
   {\bf B. Divide method}
 \end{center}

 In the case of the block BOP \cite{Ozaki} and FOE methods
 \cite{Goedecker,Stephan}, it is required to evaluate
 the modified Hamiltonian $H'=S^{-1}H$ rather than the inverse of
 overlap matrix. In such cases we have an alternative way
 where a linear equation 
 \begin{eqnarray}
   SH'=H
 \end{eqnarray}
 is solved instead of calculating the inverse.
 In conventional ways of solving the linear equation for a total system,
 the computational cost scales as the third
 power of the number of atoms $N$, while the scaling could be 
 reduced to ${\rm O}(N^2)$, making use of the sparseness
 of the overlap matrix. Therefore, Gibson et al. have proposed
 a solution of Eq.~(10) with the cutoff radii of $H$ and $S$ \cite{Gibson}.
 The linear equation Eq.~(10) can be decomposed into $N$ subspace
 linear equations for $N$ finite clusters under this constraint. 
 One solves each of the subspace linear equations for the finite clusters
 centered on atom $i$ using a conventional method such as the 
 Cholesky factorization, which results in O($N$) operations
 for the computational effort.
 However, the divide method has redundancy in the calculation
 that one has to evaluate all matrix elements of the modified
 Hamiltonian $H'$ for each finite cluster compared to the other
 ${\rm O}(N)$ inverting methods in which the elements in the inverse
 of the overlap matrix are not doubly calculated.
 Thus, the prefactor of the ${\rm O}(N)$ operations could be 
 very large for highly coordinated structures such as fcc.
 The magnitude of the prefactor will be discussed in Sec.~III.
 An iterative scheme such as the Gauss-Siedel method \cite{Jones,Foulkes}
 which is commonly used for large-scale systems is also available for
 solving the linear equation Eq.~(10). However, it has been
 recognized that the iterative scheme is computationally expensive
 \cite{Gibson}, so that the iterative scheme was
 not investigated in this study. 
 We used the logical truncation method to construct the subspace
 linear equation as well as the recursion method in the test calculations
 discussed in Sec.~III in order to compare the computational performance.

 \begin{center}
   {\bf C. Taylor expansion method}
 \end{center}

 Mauri et al. have proposed to approximate the inverse of the overlap
 matrix using the Taylor expansion in their ${\rm O}(N)$ unconstrained
 minimization method \cite{Mauri}. The overlap matrix $S$ is expressed as a
 sum of the identity ${\rm I}$ and an $O$-matrix $O$ which is the overlap
 matrix between the different orbitals:
 \begin{eqnarray}
    S = {\rm I} + O,
 \end{eqnarray}
 then we can expand the inverse of $S$ in respect to the $O$-matrix
 as follows:
 \begin{eqnarray}
    \nonumber
    S^{-1} & = & \sum_{n=0}^{\infty}(-1)^n O^n\\
           & = & {\rm I} - O + O^2 - O^3 +  \dots
 \end{eqnarray}
 The computational accuracy and efficiency of the approximation 
 by the Taylor series depend on the convergence for the summation
 of Eq.~(12). The summation in Eq.~(12) does not converge, but
 diverges, when the spectrum radius of the $O$-matrix exceeds 1.0.
 Even if the $O$-matrix has no eigenvalues which are and below -1.0,
 indicating the basis set is linearly independent, the eigenvalues
 of the $O$-matrix exceed 1.0 in most cases as shown in Sec.~III.
 In such cases, the Taylor expansion method cannot be applied.
 The matrix $O^n$ is calculated as the product of the perfect
 but highly sparse $O$-matrix, and $O^{n-1}$ with the cutoff
 radii for the elements, so that the summation to a finite order
 in Eq.~(12) can be performed with ${\rm O}(N)$ operations.

 \begin{center}
   {\bf D. Hotelling's method}
 \end{center}

 Palser and Manolopoulos \cite{Palser} have suggested evaluating
 the inverse $S^{-1}$ using Hotelling's method \cite{Recipes,Pan}.
 The method has an iterative algorithm very similar to the purification
 algorithm \cite{Palser} in the DM method.
 The convergence rate in Hotelling's method is also quadratic
 as with the DM method.
 The purification of an approximate inverse is achieved using the
 following iterative relation:
 \begin{eqnarray}
    S^{-1}_{n+1} = 2 S^{-1}_{n} - S^{-1}_{n}SS^{-1}_{n}.
 \end{eqnarray}
 In case of $S^{-1}_0 = {\rm I}$, Hotelling's method is equivalent to 
 the Taylor expansion method to a finite order described in the previous
 subsection (C). It is easy to verify that $S_1$ and $S_2$ are 
 the Taylor series to the first and third orders of the $O$-matrix,
 respectively:
 \begin{eqnarray}
    \nonumber
    S^{-1}_{1} & = & 2 S^{-1}_{0} - S^{-1}_{0}SS^{-1}_{0}\\
               & = & {\rm I} - O,
 \end{eqnarray}
 \begin{eqnarray}
    \nonumber
    S^{-1}_{2} & = & 2 S^{-1}_{1} - S^{-1}_{1}SS^{-1}_{1}\\
               & = & {\rm I} - O + O^2 - O^3.
 \end{eqnarray}
 From Eqs.~(14) and (15), we see that Hotelling's method converges
 quadratically compared to the linear convergence of Taylor
 expansion method. Thus, if Eq.~(12) is a convergent series,
 Hotelling's method should be more efficient rather than
 the Taylor expansion method.
 When the spectrum radius of the $O$-matrix exceeds 1.0,
 the identity ${\rm I}$ cannot be used as the initial guess for
 the inverse $S^{-1}$. In such cases, although it is very difficult
 to estimate a good initial matrix $S_{0}^{-1}$ for the iteration Eq.~(13),
 in this study, we use the overlap $S$ with a small prefactor $\sigma$
 derived by Pan and Reif \cite{Pan} as the initial guess:
 \begin{eqnarray}
   S^{-1}_{0} = \sigma S
 \end{eqnarray}
 with  
 \begin{eqnarray}
    \sigma = \frac{1}
            {\left(
             \displaystyle{\max_{i\alpha}}
             \displaystyle{\sum_{j\beta}}\vert S_{i\alpha,j\beta}\vert
             \right)^2}.
 \end{eqnarray}
 It is noted that Hotelling's method possesses an advantage 
 that the inverse at the previous MD step could be a good guess
 of $S_{0}^{-1}$ at the current MD step, while any information
 at the previous MD step cannot be made use of in the other methods;
 the recursion method, the divide method, and the Taylor expansion method. 
 In the iteration Eq.~(13), the elements of $S_{n}^{-1}$ are cut 
 at a finite distance. As a result of this truncation, the computational
 effort of Hotelling's method scales linearly with the system size.
 In test calculations of Sec.~III, we used the logical truncation
 method for the cutoff of the elements as in the other inverting
 O($N$) methods.

 \begin{center}
   {\bf III.~CONVERGENCE PROPERTIES}
 \end{center}

 \begin{center}
   {\bf A. Error analysis}
 \end{center}

 In order to compare the four ${\rm O}(N)$ inverse methods presented
 in the Sec.~II in terms of computational accuracy and efficiency,  
 we first relate the 1-norm of an error matrix $E$ with the error of
 eigenvalues $\epsilon_{\nu}$ of a secular equation by using
 an error analysis theory \cite{Golub,Chatelin}.
 The generalized secular equation with the overlap matrix $S$ is derived
 from the variational principle within DFT using a non-orthogonal basis set.
 \begin{eqnarray}
   S^{-1}HC_{\nu} = \epsilon_{\nu} C_{\nu},
 \end{eqnarray}
 where $H_{i\alpha,j\beta}
        \equiv \langle i\alpha \vert\hat{H}\vert j\beta\rangle$
 and
 $C_{i\alpha,\nu}$ is an expansion coefficient
 $C_{i\alpha,\nu}\equiv \langle i\alpha\vert \phi_{\nu}\rangle$
 in a one-particle wave function $\vert \phi_{\nu}\rangle$.
 Let us consider substituting the exact inverse $S^{-1}$ with
 an approximate inverse $S'^{-1}$ in Eq.~(18), then the difference
 between $S^{-1}$ and $S'^{-1}$ is 
 \begin{eqnarray}
   S'^{-1} - S^{-1} = \Delta S^{-1}.
 \end{eqnarray}
 For the approximate inverse $S'^{-1}$
 the secular equation $S'^{-1}HC'_{\nu} = \epsilon'_{\nu} C'_{\nu}$
 is satisfied with approximate eigenvalues $\epsilon'_{\nu}$ and
 eigenvectors $C'_{\nu}$.
 According to the error analysis theory \cite{Golub,Chatelin},
 the difference between the exact and the approximate eigenvalues
 is given by
 \begin{eqnarray}
   \vert\epsilon'_{\nu} - \epsilon_{\nu}\vert = {\rm O}(\lambda)
 \end{eqnarray}
 with $\lambda$, which is the 1-norm of a matrix $\Delta S^{-1}H$,
 defined by 
 \begin{eqnarray}
   \lambda  = 
            \max_{j\beta}\sum_{i\alpha}
            \left\vert \sum_{k\gamma}
            \Delta S^{-1}_{i\alpha,k\gamma}H_{k\gamma,j\beta}
            \right\vert.
  \end{eqnarray}
 Therefore, we see that the error in eigenvalue is proportional to
 the 1-norm of $\Delta S^{-1}H$ for the approximation of the overlap
 matrix. Equation (20) apparently connects the error of the overlap
 matrix to that of the eigenvalue. However, it is not possible 
 to calculate the exact inverse for infinite or periodic systems, 
 so that we introduce an error matrix $E$, which is easily evaluated,
 defined as the difference between a matrix $SS'^{-1}H$ and the original
 Hamiltonian $H$:
 \begin{eqnarray}
    \nonumber
    E & \equiv & 
         SS'^{-1}H - H\\
      & = &  S\Delta S^{-1}H.
 \end{eqnarray}
 The 1-norm $\eta$ of the error matrix $E$ can be related to
 that $\lambda$ of the matrix $\Delta S^{-1}H$ as follows:
  \begin{eqnarray}
     \nonumber
     \eta & = & \max_{j\beta}\sum_{k'\gamma'}
                \left\vert \sum_{i\alpha}\sum_{k\gamma}
                S_{k'\gamma',i\alpha}\Delta S^{-1}_{i\alpha,k\gamma}
                H_{k\gamma,j\beta}
                \right\vert\\
     \nonumber
          & \leq & 
                \max_{j\beta}\sum_{k'\gamma'}
                \sum_{i\alpha}
                \vert S_{k'\gamma',i\alpha} \vert
                \left\vert
                \sum_{k\gamma}
                \Delta S^{-1}_{i\alpha,k\gamma}
                H_{k\gamma,j\beta}
                \right\vert\\
     \nonumber
          & \leq &
                N_{av} 
                \left(
                \max_{j\beta} 
                \sum_{i\alpha}
                \left\vert
                \sum_{k\gamma}
                \Delta S^{-1}_{i\alpha,k\gamma}
                H_{k\gamma,j\beta}
                \right\vert
                \right)\\
          & = &
                N_{av}\lambda,
  \end{eqnarray}
 where $N_{av}$ is the average number of the non-zero elements
 in the overlap matrix for an orbital $\vert i\alpha \rangle$.
 The third relation in Eq.~(23) is derived by substituting the non-zero
 overlap integrals $\vert S_{k'\gamma,i\alpha} \vert$ to 1 with
 the variables $i\alpha$ fixed in the second relation.
 Considering Eqs.~(21) and (23), we can relate the 1-norm of the
 error matrix to the error of the eigenvalue:
 \begin{eqnarray}
   \vert\epsilon'_{\nu} - \epsilon_{\nu}\vert = {\rm O}(\eta). 
 \end{eqnarray}
 Therefore, we will compare the four O($N$) inverse methods
 using the 1-norm $\eta$, which is easily evaluated, instead of 
 $\lambda$.

 \begin{center}
   {\bf B. Numerical tests}
 \end{center}

 We numerically studied convergence properties of the four inverse
 ${\rm O}(N)$ methods using 1-norm $\eta$ for diamond and fcc
 Al within DFT proposed by Sankey and Niklewski \cite{Sankey}.
 In this DFT calculations we used numerical localized orbitals, 
 fireball bases by Sankey and Niklewski \cite{Sankey},
 as a minimal basis set for valence electrons.
 The radii of the radial-wave function confinement are 2.1 and 
 3.7~\AA~for carbon and aluminum atoms, respectively. 
 The minimal basis sets give 1.253 (1.244)
 and 2.515 (2.466)~\AA~ as an equilibrium bond length of dimer for
 carbon and aluminum, respectively, where the values in the
 parentheses are experimental results.

 In Fig.~(1) we show the density of states for eigenvalues of
 $O$-matrix, which is defined by Eq.~(11), in diamond and fcc Al.
 In both cases the $O$-matrices have no eigenvalues smaller than
 -1.0, so that the basis sets are linearly independent for
 the structures. However, the density of states possess finite
 values for the eigenvalues larger than or equal to 1.0 in both cases.
 In other words the spectrum radius of the $O$-matrix exceeds 1.0.
 This means that the summation in Eq.~(12) for the Taylor expansion
 method diverges for diamond and fcc Al.
 In addition to the above cases, we confirmed that the spectrum 
 radii of the $O$-matrix also exceed 1.0 for the graphite and
 poly(ethylene), so that the applicability of the Taylor expansion
 method is strictly restricted. Therefore, we do not provide the 
 convergence properties of the Taylor expansion method in this paper.

 Figure 2 shows the convergence properties of the 1-norm $\eta$ of 
 the error matrix for diamond calculated by the recursion, divide,
 and Hotelling's methods.
 In the recursion method the 1-norm exponentially decays for each
 shell cluster as a function of the number of recursion levels,
 and finally converges to the value of the 1-norm calculated by
 the divide method for the corresponding cluster.
 In the divide method the 1-norm almost exponentially diminishes
 as a function of number of shells.
 For the seven-shell cluster the 1-norm is only $3.1\times 10^{-5}$~eV.
 The identity matrix ${\rm I}$ cannot be used as an initial guess
 $S_{0}^{-1}$ in Hotelling's method because the spectrum
 radii of the $O$-matrix exceed 1.0. Thus, we gave the initial guess
 $S_{0}^{-1}$ by Eq.~(16), where $\sigma$ is 0.021 for diamond.
 In Hotelling's method the convergence properties are not 
 monotonic compared to the other two methods.
 For three-, five-, and seven-shell clusters, the 1-norm is
 gradually reduced for smaller number of iterations.
 However, the 1-norm increases after reaching at the minimum, 
 and finally we have a numerical instability that the 1-norm diverges
 as iteration proceeds.
 The smallest 1-norm for each shell-cluster is slightly larger than  
 that calculated by the divide method for the same cluster.
 Therefore, we see that Hotelling's method cannot reach the perfect 
 convergence for diamond due to the numerical instability.
 For Hotelling's method we also examined the convergence properties
 of the 1-norm $\eta$ for carbon in the diamond structure
 with 3.9~\AA~of a lattice constant in which the spectrum radius
 of the $O$-matrix is within 1.0, while the result is not shown in
 this paper. In this system the 1-norm very quickly converges to
 the corresponding value calculated by the divide method for the
 same cluster. Thus, we heuristically find that Hotelling's method
 gives convergent results for systems with the spectrum radii smaller
 than 1.0.

 As with Fig.~2, the convergence properties of the 1-norm are shown
 in Fig.~3 for fcc Al.
 The magnitude of the 1-norm is 1$\sim$2 order larger than that of
 diamond, while the behavior of the 1-norm is very similar to
 that of diamond.
 In the recursion method the converged values of the 1-norm are
 consistent with those of the divide method for four- and six-shell clusters,
 respectively. In Hotelling's method we used Eq.~(16) with $\sigma=0.0098$
 as $S_{0}^{-1}$, since the spectrum radius of the $O$-matrix exceed 1.0
 for fcc Al. The 1-norms for the four- and six-shell clusters 
 finally diverge without achieving the full convergence like for diamond.
 Although we tested the convergence properties using several values
 for $\sigma$ in both diamond and fcc Al, we could not obtain
 converged results and moreover could not avoid the numerical instability.

 Figures 4(a) and 4(b) show the relation between the magnitude of
 the 1-norm $\eta$ of the error matrix and the computational time per
 atom to evaluate the inverse of the overlap matrix for
 diamond and fcc Al, respectively.
 The comparison clearly indicates that the computational efficiency
 increases in the order of the divide $<$ Hotelling's $<$
 the recursion methods for both diamond and fcc Al.
 The recursion method is about one-hundred times faster than the divide
 method in computational time to achieve the same convergence for diamond
 and fcc Al.

 \begin{center}
   {\bf IV.~CONCLUSIONS}
 \end{center}
 
 We presented a new O($N$) algorithm for calculating the inverse of the 
 overlap matrix $S$. It is based on the recursion method with the block
 Lanczos algorithm. The problem of evaluating $S^{-1}$ is mapped to the
 block BOP method for an orthogonal TB model just by replacing the 
 Hamiltonian with the overlap operator.
 In addition, we briefly described the other known-methods
 for calculating the inverse in ${\rm O}(N)$ operations:
 the divide, the Taylor expansion, and Hotelling's methods.
 We examined the computational accuracy and efficiency
 of these ${\rm O}(N)$ inverting methods using the 1-norm of the 
 error matrix for diamond and fcc Al in DFT calculations with
 the minimal basis set for valence electrons.
 The spectrum radius of the $O$-matrix given by $(S-{\rm I})$
 exceeds 1.0 for many real materials in the DFT calculations
 based on the localized bases, which means that the applicability
 of the Taylor expansion method is significantly restricted.
 In the recursion method the 1-norm of the error matrix exponentially
 converges to the value calculated by the divide method for the same
 cluster in both diamond and fcc Al with numerical stability.
 On the other hand, Hotelling's method cannot reach the
 converged results due to the numerical instability in both cases.
 The comparison of computational time shows that the recursion 
 method is the most efficient algorithm among the four O($N$)
 inverting methods in diamond and fcc Al. The recursion method is
 about one-hundred times faster than the divide method.
 Thus, the new method for the evaluation of the inverse is
 a practical algorithm and can be incorporated
 in several O($N$) methods for total energy calculations using 
 localized orbital basis.

 \begin{center}
   {\bf ACKNOWLEDGMENS}
 \end{center}

 We would like to thank Y. Morikawa and H. Kino for helpful suggestions
 about the DFT calculations.
 We would like to thank D. R. Bowler for useful suggestions about
 ${\rm O}(N)$ inverting methods.
 Part of the computation in this work has been done using the computational
 facilities of the Japan Advanced Institute of Science and Technology (JAIST).

%

 
 \begin{figure}[t]
  \caption{\small
   The density of states for eigenvalues of the $O$-matrix for diamond
   and fcc Al, where carbon and aluminum atoms have minimal numerical
   basis sets for valence electrons which were obtained by DFT calculations
   for the atomic states. The experimental values, 3.57 and 4.05~\AA, were
   used as the lattice constants of diamond and fcc Al, respectively.}
 \end{figure}


 \begin{figure}[t]
  \caption{\small
   The 1-norm of the error matrix for diamond calculated by
   the (a) recursion, (b) divide, and (c) Hotelling's methods.
   In both the recursion and Hotelling's methods, the 1-norms were
   calculated for three-, five-, and seven-shell clusters as a function
   of number of recursion levels and iterations, respectively.}
 \end{figure}


 \begin{figure}[t]
  \caption{\small
   The 1-norm of the error matrix for fcc Al calculated by
   the (a) recursion, (b) divide, and (c) Hotelling's methods.
   In both the recursion and Hotelling's methods, the 1-norms were
   calculated for four- and six-shell clusters as a function
   of number of recursion levels and iterations, respectively.
   }
 \end{figure}


 \begin{figure}[t]
  \caption{\small
   The 1-norm of the error matrix for (a) diamond and (b) fcc Al
   against the computational time taken per atom calculated by three
   O($N$) inverting methods. The calculations were performed using
   single processor on a compaq ES40 workstation.}
 \end{figure}

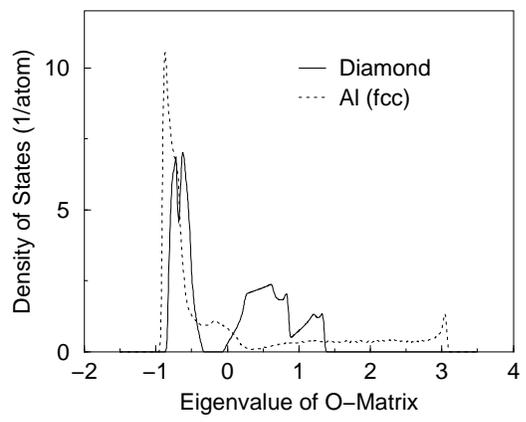
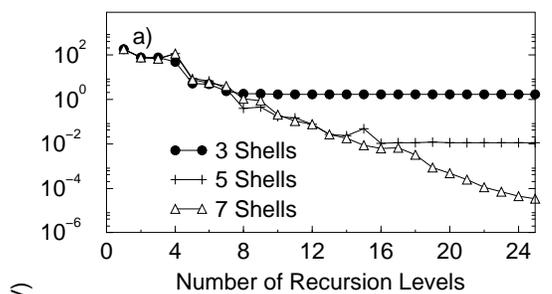
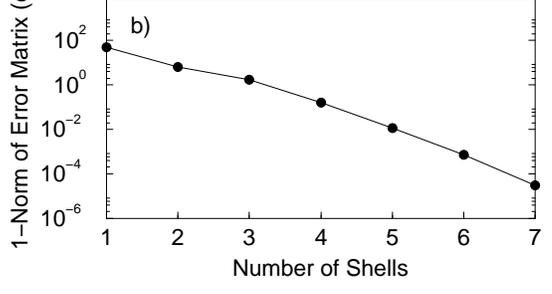
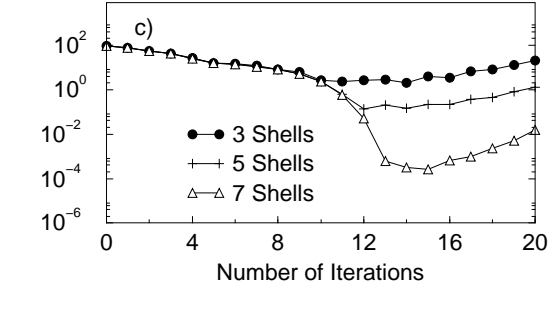
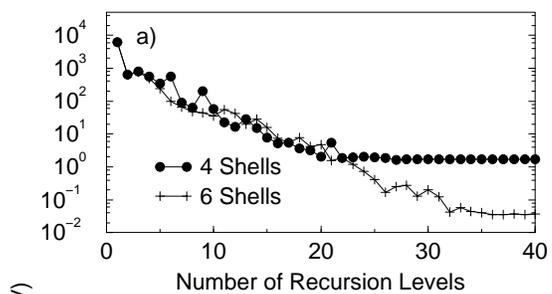
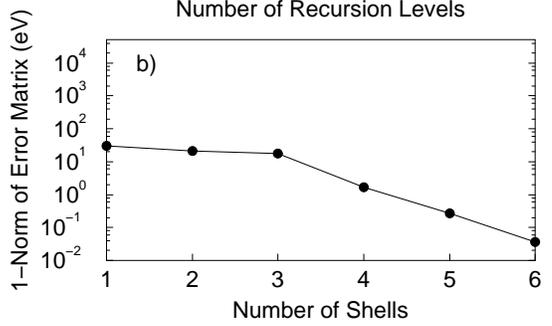
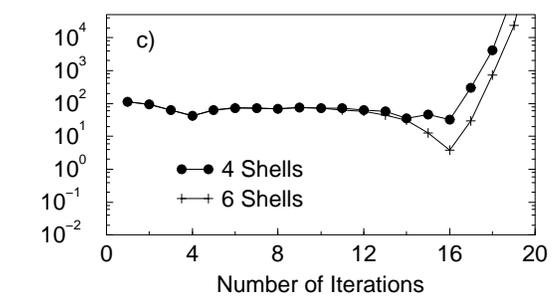
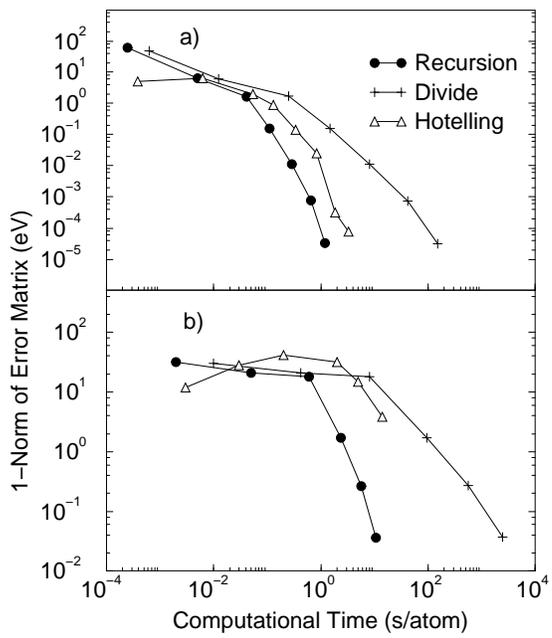
\end{document}